# Parameter free quantitative analysis of atom probe data by correlation functions: application to the precipitation in Al-Zn-Mg-Cu

Huan Zhao[1], Baptiste Gault[1], Dirk Ponge[1], Dierk Raabe[1], Frédéric De Geuser[2,*]

[1] Max-Planck-Institut für Eisenforschung, Max-Planck-Str. 1, 40237 Düsseldorf, Germany

[2] Univ. Grenoble Alpes, CNRS, Grenoble INP, SIMaP, F-38000 Grenoble, France

**Abstract**:
Atom probe tomography enables precise quantification of the composition of second phase particles from their early stages, leading to improved understanding of the thermodynamic and kinetic mechanisms of phase formation and quantify structure-property relationships. Here we demonstrate how approaches developed for small-angle scattering can be adapted to atom probe tomography. By exploiting nearest-neighbor distributions and radial distribution function, we introduce a parameter free methodology to efficiently extract information such as particle size, composition, volume fraction, number density and inter-particle distance. We demonstrate the strength of this approach in the analysis of a precipitation-hardened model Al-Zn-Mg-Cu high-strength lightweight alloy.

Al-Zn-Mg-(Cu) alloys of the 7XXX Al-alloys are precipitation-strengthened Al-alloys exhibiting high-strength and expected to play an important role for lightweight transportation. The precipitation sequence in the bulk of Al-Zn-Mg-Cu alloys has been studied and is well established [1–3]: solid solution → GP zones→ metastable η'→ stable η ($MgZn_2$). Atom probe tomography (APT) provides three-dimensional (3D), nanoscale analytical mapping of materials, and is uniquely positioned to measure the local chemical composition and 3D morphology of individual precipitates. However, early designs of the instruments had an angular field-of-view limited to less than 10 degrees and thus only small particle numbers would typically be intercepted and analyzed [4], thereby limiting the possibility of deriving precisely important metallurgical measures e.g. precipitate size, particle spacing, composition, volume fraction, and number density. The significant increase in the field of view of the technique associated with modern microscope designs [5] often allow for hundreds or thousands of nanoparticles to be captured within a single dataset [6], which makes data extraction and processing tedious, and most often erroneous [7].

Over 50 years of theoretical developments in the small-angle scattering community have enabled the development of a complete formalism to fit experimental results with analytical functions to derive microstructural characteristics [8,9]. Here, we introduce a new methodology to process radial distribution function (RDF)-based model fitting from atom probe data to provide complete characterization of both the precipitates and the matrix, without the need for any input parameters. Parameter-selection by various users is a key hurdle in ensuring reliability and reproducibility of data analysis as [10–13]. Our methodology makes use of widely available integral metrics, the distribution of distance to the first nearest-neighbor (1NN) and species-specific RDF computed on the entire dataset, making use of all available information, and thereby improving statistics. We evidence the strength of this new approach through the analysis of the temporal evolution of the precipitates formed during aging of a model Al-Zn-Mg-Cu alloy.

An Al-Zn-Mg-Cu alloy was cast into a 200mm*190mm*40mm ingot in a vacuum induction furnace. The ingot was homogenized at 475 °C for 1.5 hours and water quenched followed by hot rolling at 450 °C to 3mm thickness and water quenched. Samples were then solution treated at 475 °C for an hour and quenched in water, immediately followed by an aging treatment at 120ºC for 30 minutes, 2 h and 24 h. An overaging step was performed on the 24h aged sample by further aging for 6 h at 180°C. A specimen was also analyzed immediately after quench. It will be further called 'as quenched' even though it has in fact undergone roughly 6h of natural ageing before APT analysis which may have already caused some degree of clustering. Table 1 summarizes the chemical composition measured by wet chemical analysis after homogenization.

Table 1. Bulk chemical composition of a model Al-Zn-Mg-Cu alloy (in at%)

| Alloy  | Zn   | Mg   | Cu   | Zr   | Fe   | Si    | Al      |
|--------|------|------|------|------|------|-------|---------|
| (at %) | 2.69 | 2.87 | 0.95 | 0.05 | 0.01 | <0.01 | balance |

Specimens for APT measurement and analysis were prepared to investigate the precipitation in the bulk by using a two-stage electrochemical polishing protocol described in [14]. APT was performed in high-voltage pulsing mode with a pulse fraction of 20% at a repetition rate of 250 kHz and at a base temperature of 50K in a Cameca Local Electrode Atom Probe (LEAP) 5000XS. The total voltage was increased to maintain an average detection rate of 1 evaporative event per 100 pulses. Datasets containing 50-150 million ions were acquired for each aging state. Data reconstruction and analysis were performed using the Integrated Visualization and Analysis software (IVAS 3.6.14). The partial structural information from within the dataset allowed to calibrate the reconstructions parameters as described by Gault et al. [15].

The aging response was followed by Vickers hardness measurements using a 0.5 kg load. The values reported in Figure 1 are the average of 10 individual measurements. The hardness of the supersaturated solid solution state is 120±3 HV0.5. The hardness then progressively increases up to 24 hours, with a peak hardness of 185±5 HV0.5. The overaging step resulted in significant coarsening and a decrease in hardness down to 171±3 HV0.5. Representative datasets across the aging process are shown inset in Figure 1. Iso-surfaces encompassing regions in the reconstruction containing over 10at% Zn highlights the formation and coarsening of Zn-Mg-rich precipitates.

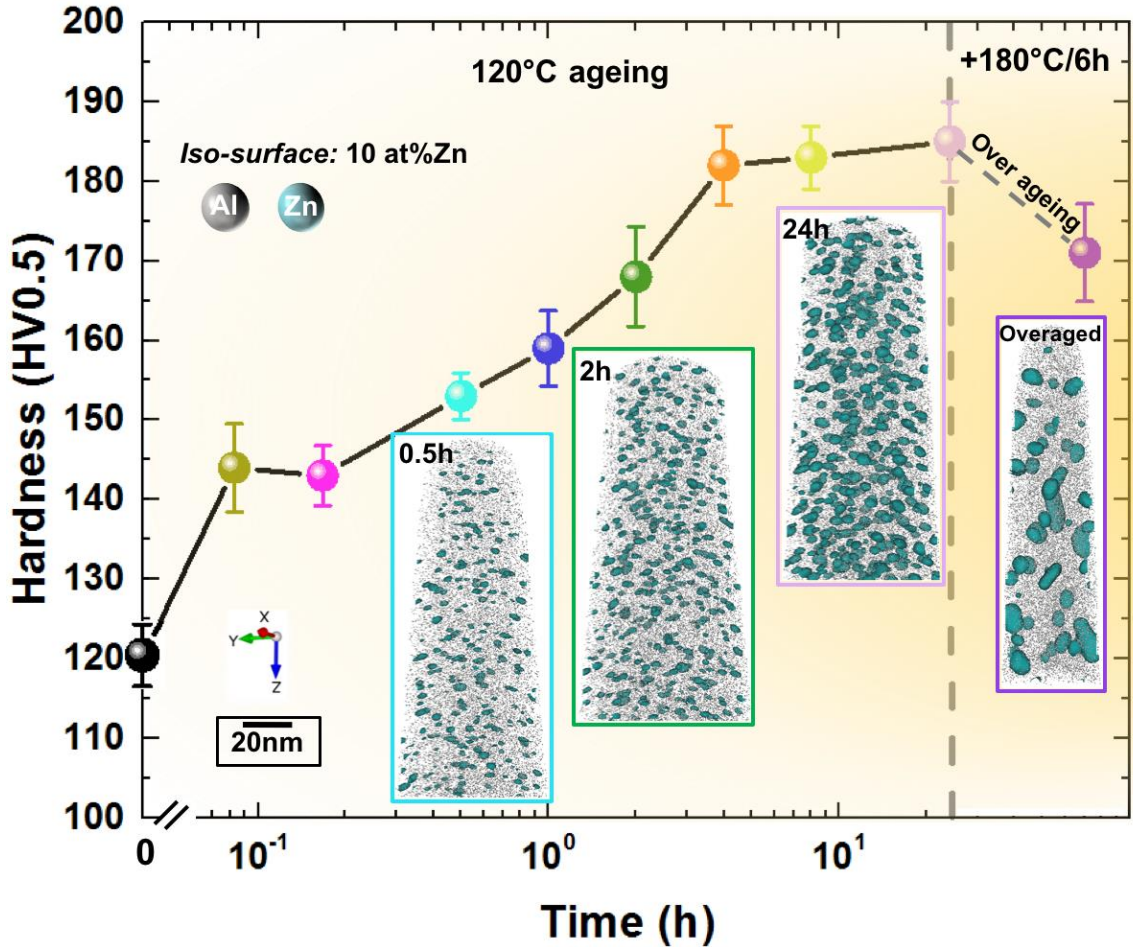

*Figure 1:* Evolution of the hardness as a function of aging time. Inset are atom maps to show the effect of nanostructures on the hardness. Al atoms are shown in grey and Zn atoms in dark cyan. The precipitates are visualized with 10 at% Zn iso-composition surface. Only part of the atoms maps are shown in order to obtain the same scale for clarity. The atom maps are to scale.

From a typical dataset, two integral information can be obtained: the global composition and the matrix composition. The former is derived from counting the relative number of ions of each element within the mass spectrum. The latter can be derived from the distribution of isolated atoms for the determination of the matrix composition (DIAM) method [16]. DIAM is based on the probability for each solute species to have no neighbor (i.e. to be isolated) at a given distance $r$, in comparison to that obtained from a random distribution of solutes. Both can be directly derived from the first nearest-neighbor distribution.

For characterizing the precipitates present in the volume, we use radial distribution functions calculated with each solute of a specific species as the central atom. The calculation is then repeated for each species separately. The RDF $n_{i-j}(r)$ is the number of atoms of species $j$ at a distance $r$ of an atom of species $i$. To speed up the calculation, the RDF is computed on a randomly-selected subset of atoms, which does not affect the results, since the RDF is normalized by the composition. Calculation of an RDF for one solute in a 40 million ions dataset was timed to take approx. 20 minutes, while it lasted only approx. 1 minute on a subsampled dataset (1 in 40 ions). Even though subsampling increases statistical noise for low counts and should be used with care, no significant effect was observed.

The $n_{i-j}(r)$ enable to compute the average composition $C_{i-j}(r)$ of element $j$ at a distance $r$ of an element $i$:

$$C_{i-j}(r) = \frac{n_{i-j}(r)}{\sum_j n_{i-j}(r)} \quad (1)$$

We call $\overline{C_i}$ the average composition of element $i$. In an isotropic system, $\overline{C_i} \cdot C_{i-j}(r)$ is the correlation (convolution) of the spatial composition of element $i$ and that of element $j$.

$$\overline{C_i} \cdot C_{i-j}(r) = <C_i(\vec{r_1}) \cdot C_j(\vec{r_2})> \text{ with } r = |\vec{r_2} - \vec{r_1}| \quad (2)$$

At large $r$ this convolution necessarily approaches $\overline{C_i} \cdot \overline{C_j}$. We can subtract this final value and introduce the $i - j$ pair correlation function (PCF) $\gamma_{i-j}(r)$:

$$\gamma_{i-j}(r) = \overline{C_i} \cdot C_{i-j}(r) - \overline{C_i} \cdot \overline{C_j} \quad (3)$$

Where $\gamma_{i-j}(r) = 0$ when there is no correlation, which is always true for large $r$. It can be re-written:

$$\gamma_{i-j}(r) = <\left(C_i(\vec{r_1}) - \overline{C_i}\right) \cdot \left(C_j(\vec{r_2}) - \overline{C_j}\right)> \text{ with } r = |\vec{r_2} - \vec{r_1}| \quad (4)$$

or

$$\gamma_{i-j}(r) = <\Delta C_i(\vec{r_1})\Delta C_j(\vec{r_2})> \text{ with } r = |\vec{r_2} - \vec{r_1}| \quad (5)$$

to highlight that only the fluctuation of composition $\Delta C_i$, i.e. the difference with the average, is relevant to the PCF. This correlation function is analogous to that introduced by Debye and Bueche [17] and which the basis of the small-angle scattering formalism. We can further introduce a normalized correlation function $\gamma^0_{i-j}(r)$ by writing:

$$\gamma_{i-j}(r) = \overline{\Delta C_i \Delta C_j} \cdot \gamma^0_{i-j}(r) \quad (6)$$

where $\gamma^0_{i-j}(0) = 1$ and $\gamma^0_{i-j} = 0$ at large $r$. $\gamma^0_{i-j}(r)$ is only dependant on the geometry of the objects. In particular, for $i = j$, we have the auto-correlation:

$$\gamma_{i-i}(r) = \overline{\Delta C_i^2} \cdot \gamma^0_{i-i}(r) \quad (7)$$

From its definition, the value of $\gamma_{i-j}$ at the origin $\overline{\Delta C_i^2}$ is the mean squared composition fluctuation, which directly depends on the quantity and composition of the objects. Contrasting with $\gamma^0_{i-i}(r)$, $\overline{\Delta C_i^2}$ is exclusively composition-based, it is essentially immune to typical APT reconstruction artefacts such as local magnification or preferential field evaporation caused by the difference in the electric field necessary to induce field evaporation for different species or phases [18,19]. In small-angle scattering, the separation of the geometry-based normalized correlation function and the mean squared fluctuation enables the measurement of the size of objects irrespective of their quantity or of the scaling of the measurement. In APT, especially for very small objects, the composition is likely more accurate, while $\gamma^0_{i-j}$ is likely more questionable.

We assume 2 phases of homogeneous composition in element $i$, i.e. precipitates composition $C^i_p$ in a matrix of composition $C^i_m$. We further assume that both share the same atomic volume $\Omega$. While this is never strictly true, the difference in atomic volumes between phases is usually small compared to the possible apparent density differences due to e.g. local magnifications [18]. In these conditions, we can write the precipitates volume fraction as:

$$f_v = \frac{\overline{C_i} - C^i_m}{C^i_p - C^i_m} \quad (8)$$

which should be true for any $i$.

Within these assumptions we can calculate the average of the product of the composition fluctuations:

$$\overline{\Delta C_i \Delta C_j} = f_v(1 - f_v)(C^i_p - C^i_m)(C^j_p - C^j_m) \quad (9)$$

In particular, for $i = j$

$$\overline{\Delta C_i^2} = f_v(1 - f_v)(C^i_p - C^i_m)^2 \quad (10)$$

or

$$\overline{\Delta C_i^2} = (C^i_p - \overline{C_i})(\overline{C_i} - C^i_m) \quad (11)$$

The value of the pair correlation function at $r = 0$, in combination with the matrix composition obtained from DIAM, provides a mean to derive both the volume fraction and the precipitates composition, irrespective of the particles shape and without being affected by local magnification.

The normalized correction function $\gamma_0(r)$ can be calculated analytically for simple shapes. For a sphere of radius $R$, i.e. of homogeneous density and with a sharp interface with the matrix, it is given by [20,21]:

$$\gamma_0^{sphere}(r, R) = \begin{cases} 1 - \frac{3r}{4R} + \frac{r^3}{16R^3} & (r \leq 2R) \\ 0 & (r > 2R) \end{cases} \quad (12)$$

For a size distribution of spheres (e.g. normal, log-normal,…) this can be readily integrated numerically. We have fitted all the experimental $\gamma_{i-j}(r)$ for all aging conditions independently, but with the same size for all $i-j$ pairs. We use a lognormal distribution of spheres with a 20% dispersity. We also add a contribution of an excluded volume [22], which is a small correction to the PCF depending on the size and volume fraction which gives a negative contribution originating from the impossibility of having two precipitates on top of each other. This contribution essentially helps to reproduce the fact that some curves go through a negative minimum before reaching zero for larger $r$. In any case, the validity of the fitting model for $\gamma_0(r)$ should be assessed in regard with the distortions due to local magnifications as well as with the validity of the chosen shape and size dispersion. Nevertheless, it provides a robust effective average size measured through all existing objects in the volume.

The results are a complete set of $\overline{\Delta C_i^2}$ and $R$ for all times and $i$. The next step is to compute the $C_p^i$ from eq. (11) and $C_p^{Al} = 1 - C_p^{Zn} - C_p^{Mg} - C_p^{Cu}$. From these $C_p$ we can derive the volume fractions $f_v$ from eq. (10) which gives us a set of volume fractions calculated for each solute element. Since we assume that there is a single type of object, we will reduce the uncertainty by fixing the volume fraction as the average of that computed from eq. (10) with Zn and Mg (the low content of Cu both on average and in the precipitates, makes the calculations noisy for this element). The mean diameter of the precipitates is obtained from the fit, but as mentioned above, is affected by possible local magnification artefacts. The number density is derived from the ratio of the volume fraction to the average volume.

This combined DIAM and RDF-based protocol was applied to characterize the matrix composition and the precipitates evolution over the course of the aging process. Figure 2 (a–b) report the evolution of matrix composition as well as the composition measured in the entire dataset, as well as the precipitates composition over the course of the aging process. Figure 3 (a–d) show the volume fraction of precipitates, their average radius, the number density, and the distance between precipitates roughly estimated as $n_v^{-3}$. Since the solute content of the precipitates evolves with ageing, we chose to represent only the contribution of solutes to the volume fraction to better visualize the advancement of the reaction.

While Cu low composition may be more sensitive to noise and ranging issues, Zn and Mg global dataset composition should be essentially independent of aging time and equal to the nominal composition. While this is observed initially, the global composition in Zn, and to some extent in Mg, eventually drops to a lower value in the peak aged state, and more importantly so after over aging. This could only be attributed to a loss of solutes to grain boundaries or to often-reported instrumental issues related to the preferential field evaporation of specific solutes [23].

The results show a number density in the range of $10^{25}$ m$^{-3}$ of solute-rich clusters approx. 1 nm in size form in the as-quenched state. This might be caused by the natural aging during sample preparation. The precipitates composition in the three solutes plateaus after 30 min to approx. 75 at % Al, 13 at % Zn, 10 at% Mg and 2 at % Cu. Over the first 2 hours at 120°C aging, nearly spherical zones enriched in solutes form and develop into GP zones with a number density of $10^{24}$–$10^{25}$ m$^{-3}$. In the peak aged condition, the precipitates' composition does not reach that of the η' phase and thus can be considered as GP zones. It should also be noted that within the precipitates, Cu remains in low composition and almost unchanged at 120°C aging, while exhibiting a strong increase in the overaging step. This likely originates from the low diffusivity of Cu, together with the solubility in the precipitates [24]. Since its content was stable from the beginning, however, it means that the very early clusters/GP zones are comparatively richer in Cu, which may indicate that Cu plays a significant role in GP zone nucleation. Significant coarsening is evidenced in the overaged state as revealed by the increase of precipitates radius, volume fraction and inter-precipitates distance. Only then does the composition of the precipitates reach a content compatible with η'.

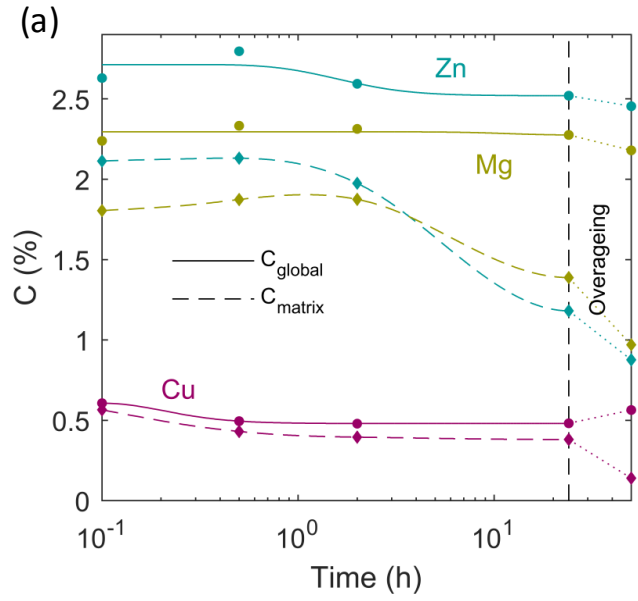

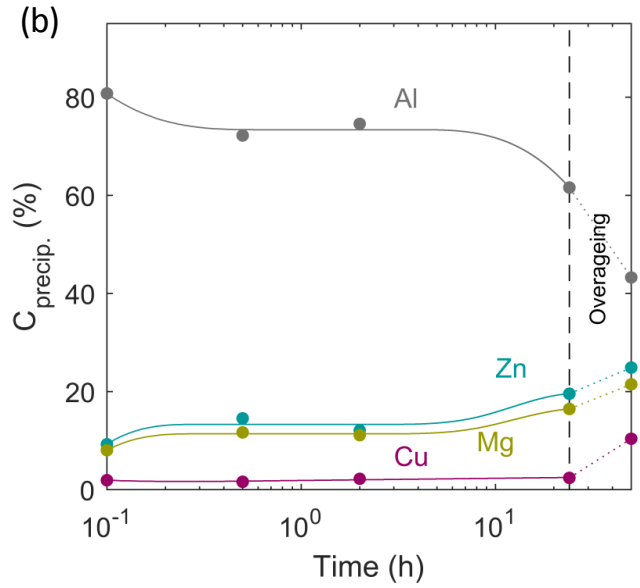

*Figure 2: (a) evolution of the composition measured in the entire datasets (discs, the solid line is a guide to the eye) and the matrix composition (diamonds, the dashed line is a guide to the eye) for the different solutes as a function of aging time. (b) evolution of the precipitates composition (discs, the dashed line is a guide to the eye) for the different solutes as a function of aging time.*

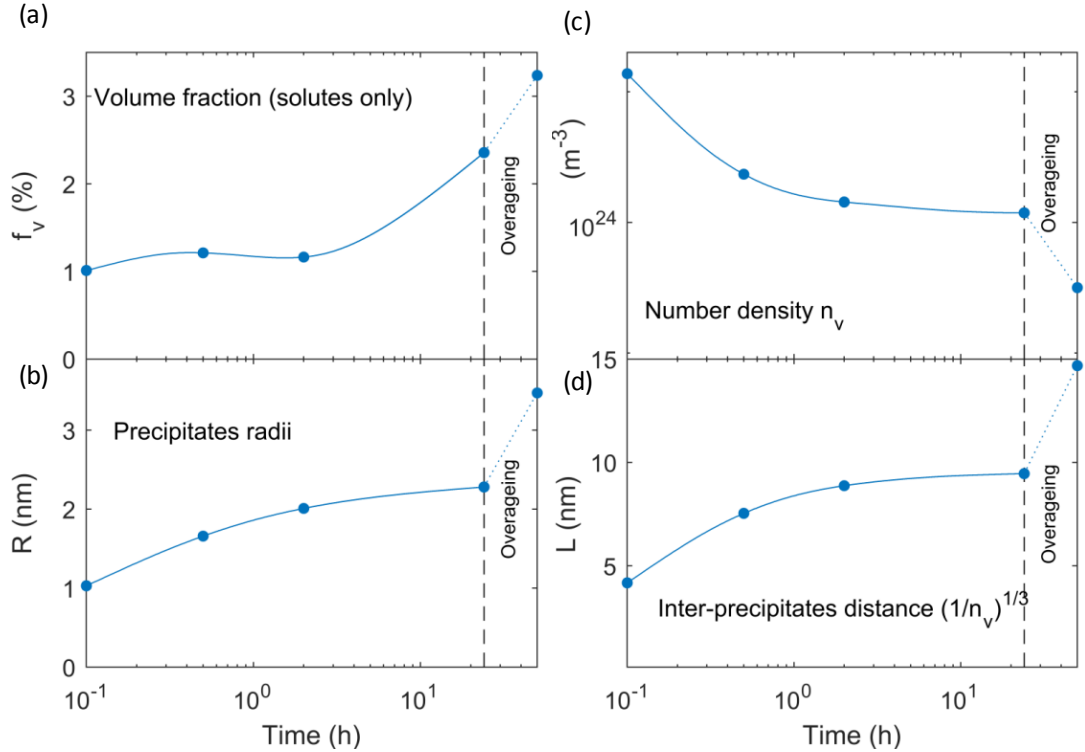

*Figure 3: Evolution during the aging process of (a), the volume fraction of precipitates; (b), the precipitate radius; (c), volume density of precipitates; (d), average distance between precipitates.*

Our approach provides all the meaningful metallurgical information, similarly to what would usually be performed in small-angle scattering studies [25,26], except that here the composition of the objects can be directly determined. This information is necessary to develop mean-field models to explain the mechanical behavior of materials [27] but is rarely available directly from APT. The methodology we introduce here is parameter-free, versatile and can be applied to numerous systems, provided that the assumptions from the model are not invalidated. In cases where a large number of objects are imaged within the dataset, our alternative approach ensures the highest possible statistical significance of the results, as well as reproducibility of the results.

## Acknowledgements

H. Zhao would like to acknowledge the Chinese Scholarship Council for the PhD scholarship granted to support this work. U. Tezins and A. Sturm are acknowledged for their support in the use of the atom probe and PFIB facility at MPIE.